\documentclass[twocolumn,amsmath,amssymb]{revtex4}
\usepackage{amsmath}
\usepackage{bm}
\usepackage{graphicx}
\usepackage{xspace}

\begin{document}

\newcommand{\bkg}{B(\bm{x})}
\newcommand{\ilm}{I(\bm{x})}
\newcommand{\obj}{\Pi(\bm{x})}
\newcommand{\sph}{\mathcal{S}(\bm{x} - \bm{x_p})}
\newcommand{\psf}{P(\bm{x}-\bm{x'}; \bm{x})}
\newcommand{\im}{\mathcal{I}(\bm{x})}
\newcommand{\xfit}{\bm{x_{f}}}
\newcommand{\grad}{\bm{\nabla}}
\newcommand{\gradf}{\bm{\nabla_f}}
\newcommand{\tp}{\textit{trackpy}}
\newcommand{\xcentroid}{\bm{x}_\mathrm{cm}}

\newcommand{\angstrom}{\AA\xspace}

\let\tempone\itemize
\let\temptwo\enditemize
\renewenvironment{itemize}{\tempone\addtolength{\itemsep}{-0.5\baselineskip}}{\temptwo}

\let\tempthree\enumerate
\let\tempfour\endenumerate
\renewenvironment{enumerate}{\tempthree\addtolength{\itemsep}{-0.5\baselineskip}}{\tempfour}

\newcommand{\nopix}{No pixelation, \textit{i.e.} access to $\im$ everywhere.}
\newcommand{\pix}{Pixelation, \textit{i.e.} access to $\im$ at discrete points.}

\newcommand{\flatilm}{Uniform illumination: $\ilm=1$.}
\newcommand{\gentleilm}{Simply-varying illumination: $\ilm = 1 + \alpha x$.}
\newcommand{\realilm}{Realistic, spatially-varying illumination, measured from a microscope image.}

\newcommand{\simplepsf}{A simple, Gaussian point-spread function:\\
        $\psf = 1 / \sqrt{2\pi\sigma^2} \times \exp(-(\bm{x}-\bm{x'})^2/2\sigma^2)$}
\newcommand{\realpsf}{A point-spread function calculated from diffraction theory and including aberrations from refractive index mismatch~\cite{Hell1993}.}

\newcommand{\onesphere}{A single sphere:\\
    $\obj=\sph$, where $\sph=1$ for $|x_p| < a$ and $0$ otherwise.}
\newcommand{\onesphereshort}{A single sphere.}
\newcommand{\twosphere}{Two spheres, with surface-to-surface separation $\delta$.}

\newcommand{\flatbkg}{Uniform, zero background: $\bkg=0$, $c=0$.}
\newcommand{\nonoise}{No noise.}

\title{Biases in particle localization algorithms}
\author{Brian D. Leahy}
\affiliation{Dept. of Physics, Cornell University}
\thanks{Currently at SEAS, Harvard University}
\author{Matthew Bierbaum}
\affiliation{Dept. of Physics, Cornell University}
\author{James Sethna}
\affiliation{Dept. of Physics, Cornell University}
\author{Itai Cohen}
\affiliation{Dept. of Physics, Cornell University}

\begin{abstract}
Automated particle locating algorithms have revolutionized microscopy image analysis, enabling researchers to rapidly locate many particles to within a few pixels in a microscope image. The vast majority of these algorithms operate through heuristic approaches inspired by computer vision, such as identifying particles with a blob detection. While rapid, these algorithms are plagued by biases~\cite{Baumgartl2005, Lu2013, Gao2009}, and many researchers still frequently ignore or understate these biases. In this paper, we examine sources of biases in particle localization. Rather than exhaustively examine all possible sources of bias, we illustrate their scale, the large number of sources, and the difficulty of correcting the biases with a heuristic method. We do this by generating a series of simple images, introducing sources of bias one at a time. Using these images, we examine the performance of two heuristic algorithms throughout the process: a centroid algorithm and a Gaussian fitting algorithm. We contrast the two heuristic methods with a new approach based on reconstructing an image with a generative model to fit the data (Parameter Extraction from Reconstructing Images, or PERI). While the heuristic approaches produce considerable biases even on unrealistically simple images, the reconstruction-based approach accurately measures particle positions even in complex, highly realistic images. We close by reiterating the fundamental reason that a reconstruction-based approach accurately extracts particle positions -- any imperfections in the fit both demonstrate which sources of systematic error are still present and provide a roadmap to incorporating them.
\end{abstract}

\maketitle

Over the past three decades, computer analysis has revolutionized microscopy. The improvement is especially salient in particle tracking experiments. Modern algorithms inspired by computer vision automatically locate thousands of particles with near-pixel or sub-pixel accuracy, all in a few blinks of an eye~\cite{Crocker1996, Andersson2008, Anthony2009, Gao2009, Parthasarathy2012, Besseling2015, Leocmach2013}. However, it has proven difficult to extend the accuracy of these particle localization algorithms below a fraction of a pixel, especially in complex images. In a recent paper, we proposed a different methodology for measuring particle positions. We advocate measuring parameters such as particle positions by reconstructing the entire image, in a method called Parameter Extraction from Reconstructing Images or PERI~\cite{Bierbaum2017}. In this letter, using a series of examples, we demonstrate many of the limitations of previous, heuristic particle localization methods, and we show how a generative model framework based on reconstructing the image (such as PERI) can overcome these limitations.

Many of the users and creators of these heuristic algorithms are aware of the presence of biases, as evidenced by the extensive literature on the subject~\cite{Cheezum2001, Baumgartl2005, Rogers2007, Jenkins2008, Gao2009, Parthasarathy2012, Lu2013, Chenouard2014, Jensen2016, Burov2017} and from many personal conversations we have had with members of the community. Some of what we say in this letter is not new. Nevertheless, we feel that the the magnitude of biases and their myriad of sources is under-appreciated by most, and we would like to use this space to emphasize and explore sources of biases in particle localization algorithms.

By a ``heuristic'' method, we mean any method that measures properties of an image (such as particle positions) with an incomplete proxy, usually without checking the proxy against the measured intensity data itself through a generated image. For example, locating particles with an intensity centroid is a heuristic method. Such an algorithm measures the intensity centroid in a region of interest, interpreting the centroid as the particle's physical position. The centroid algorithm does not reconcile the resulting position against the intensity values of the image itself, leaving no way to check whether the centroid proxy adequately describes the data. As we will show below, a proxy of an intensity centroid does not adequately describe real images. Likewise, refining particle positions by fitting a region to a Gaussian blur is a heuristic. Since a real image of a particle is quite different from a Gaussian blur, there is a considerable difference between the measured intensity data from a real image and even the best-fit Gaussian blur. As such, the Gaussian-fitting proxy is inadequate for real images, and its positions will be biased, as we show below. Similarly, locating particles by finding a center of radial symmetry~\cite{Parthasarathy2012} or by finding the center from which the image's intensity decays~\cite{Andersson2008} are also heuristics. While all these heuristic methods use the image intensity to estimate particle positions, at no point in their algorithms is the inferred data compared to the raw intensities of the image itself.

A reconstruction approach fundamentally differs from these heuristic methods. In a reconstruction approach, a model image, based on the detailed physics of image formation, is created and compared directly to the raw intensity data. Examining the difference between the model and the data provides a rigorous way to test the quality of the extracted parameters, through a Bayesian examination of the evidence. A good reconstruction, one that is based off a complete model of the image formation, will correctly describe the raw image intensities, provide physically accurate parameters, and capture all the signal in the data possible.

A detailed comparison of a reconstruction-based approach against a commonly-used centroid method is shown for a suite of realistic images in the supplemental information of ref.~\cite{Bierbaum2017}. However, looking at these plots doesn't make it clear \textit{why} the heuristic method returns imperfect results. This document explores the \textit{why} a little deeper. To that end, we'll start with the simplest possible image and gradually add complexity, seeing how the heuristic methods compare to a reconstruction-based approach (PERI) all the while. 

Throughout we compare PERI with two heuristic methods. The first method identifies particles centers through intensity centroids, as developed by Crocker and Grier~\cite{Crocker1996} and implemented in Python as \tp{}~\cite{trackpy}. The second method refines the particle positions by local fitting to a Gaussian, inspired by many algorithms such as ref.~\cite{Anthony2009}, but using our own implementation. We've picked these two heuristic methods because they are some of the most commonly used. However, the biases investigated below will apply generically to any heuristic method. 

Briefly, the centroid method as implemented in \tp{} works by smoothing the image for noise- and background-subtraction, then identifying bright spots in the image as particles. It then refines the positions of those particles by finding the centroid in a region-of-interest around the bright pixel. You can find a description of the centroid method in detail online~\cite{trackpy}. To refine the particle positions by fitting the image to a Gaussian, we fit the amplitude, center, and width of the Gaussian, as well as a nonzero offset. For the images with multiple particles, we've selected a region-of-interest around each particle and fit to a Gaussian. For images with multiple particles, we've used the initial image the size of the region-of-interest of the particle.

To analyze images with PERI, we always used a complete model to fit the images. To keep the comparisons fair and as close to real as possible, we randomize all the parameters in PERI's model before starting the fit, including the illumination and point-spread function parameters and the particle positions and radii. 

Our mathematical model for confocal image formation is given in ref.~\cite{Bierbaum2017}: a distribution of dye $\Pi$ is illuminated by a laser $I$, convolved with a point-spread function $P$, and imaged on detector with a nonzero background $B$:
\begin{equation}
\begin{split}
\mathcal{M}(\bm{x}) = & B(\bm{x}) + 
\\
& \int \, d^3\bm{x'} \left[ I(\bm{x'})(1 - (1-c)\Pi(\bm{x'})) \right] P(\bm{x} - \bm{x'}; \bm{x}) \; ,
\end{split}
\label{eqn:fullmodel}
\end{equation}
the constant offset $c$ partially captures rapid variations in the background, as described in ref.~\cite{Bierbaum2017}. Throughout, we generate images $\mathcal{I}(\bm{x})$ that takes a particular model image $\mathcal{M}(\bm{x})$ and possibly adds noise $\mathcal{N}(\bm{x})$:
\begin{equation}
\mathcal{I}(\bm{x}) = \mathcal{M}(\bm{x}) + \mathcal{N}(\bm{x}) \quad .
\end{equation}

\section{Idealized Heuristics work perfectly for Idealized Images.}

To start, we'll examine the simplest possible image:
\begin{enumerate}
\item \nopix
\item \flatilm
\item \simplepsf
\item \onesphere
\item \flatbkg
\item \nonoise
\end{enumerate}

For this simple case, the object function $\sph$ is symmetric about the particle's center. Neither the symmetric point-spread function nor the constant illumination or background break this symmetry. Thus, this idealized image is symmetric about the particle's position, as can be seen from equation~\ref{eqn:fullmodel}. By symmetry, the center-of-mass of the image will be \textit{exactly} at the particle's position. In other words, for this idealized image a centroid method works \textit{exactly}~\footnote{Even for an un-pixelized, symmetric image, the centroid method can still be biased. The center-of-mass is measured through an integral: $\xcentroid = \int \bm{x} \im \, d\bm{x} / \int \im \, dx$. The symmetry argument that the centroid is correct is only valid if the integral converges. While this may seem like a mathematical technicality, most images have a nonzero intensity far away from an object. Since $\im$ does not go to zero at large $\bm{x}$, the integral cannot converge. As a result, a non-zero background is a major cause for biases in centroid methods~\cite{Berglund2008}. Most common centroid methods, including \tp{}, adjust for this by subtracting long-wavelength variations in intensity from the image.}.

Likewise, fitting to a simple, symmetric function like a Gaussian typically gives the exact answer. The image $\im$ is even about $\bm{x_p}$. Our fitting heuristic fits the image to some function $f(\bm{x} - \xfit)$. The best fit minimizes the cost $\chi^2$:
\[
\chi^2 (\xfit) = \int_{-\infty}^\infty \left[ \im - f(\bm{x}-\xfit) \right]^2 \, d \xfit
\]
Taking a gradient with respect to the fitted position $\xfit$ gives
\begin{equation}
\bm{\nabla_{f}} \chi^2 = \int_{-\infty}^\infty  2 \left[ \im - f(\bm{x}-\xfit) \right]  \bm{\nabla_{f}} f(\bm{x}-\xfit) \, d\xfit
\label{eqn:perf_fit}
\end{equation}
There will be a local extremum in the cost $\chi^2$ when this gradient is 0. The Gaussian fitting function $f(\bm{x}-\xfit)$ is even about $0$, and its gradient is odd about $0$. Since $\im$ is even about $\bm{x}_p$, the integral in equation~\ref{eqn:perf_fit} is exactly zero when $\xfit = \bm{x}_p$, and the fit will return the correct, unbiased position.

So what's the problem with heuristic methods? The problem is that the real images that you take on your microscope are not the idealized images dreamt about above.

\section{Single Particle: Pixelation}
\label{sec:px}

\begin{figure*}
\includegraphics[width=\textwidth]{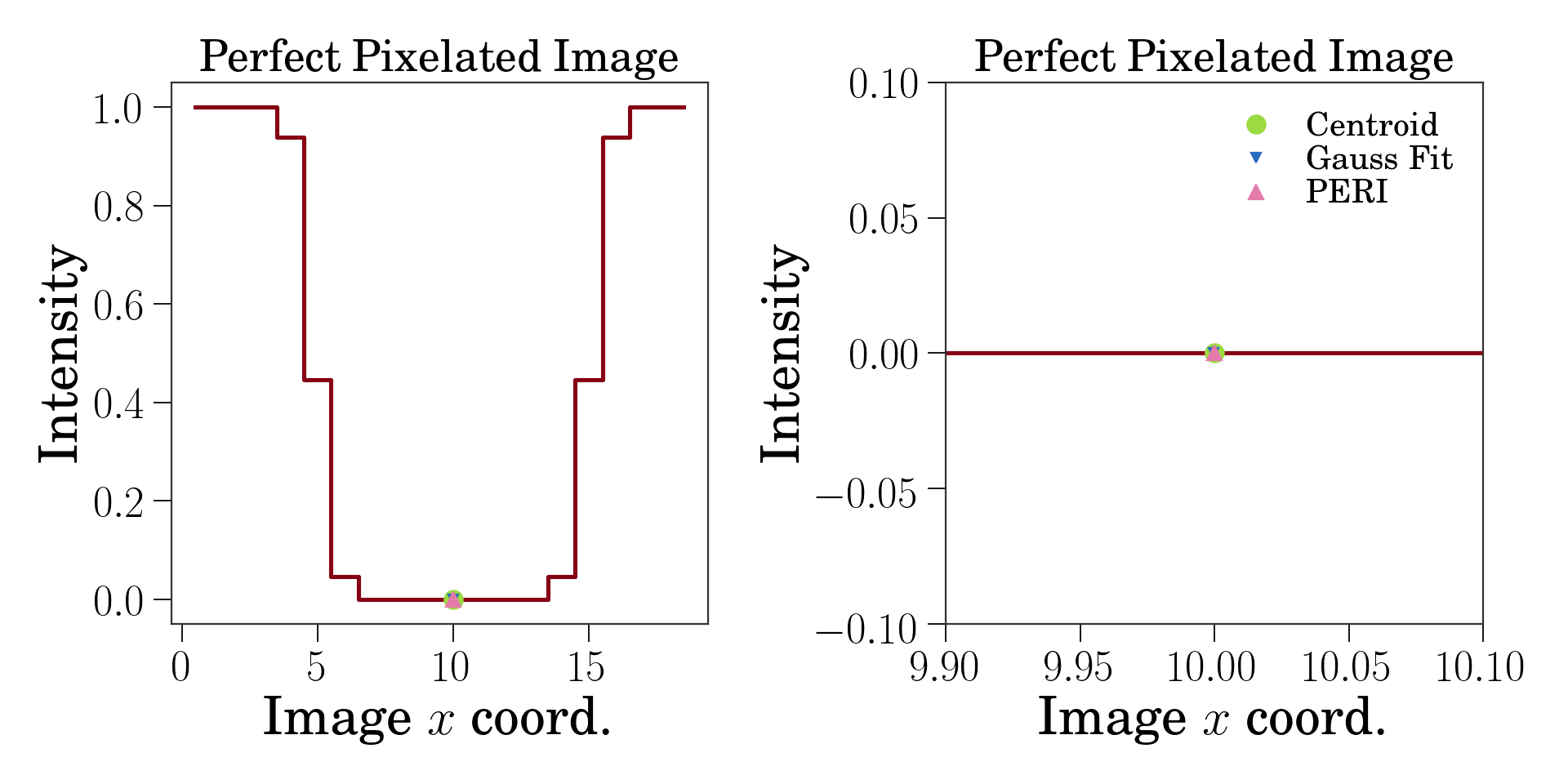}
\caption{Left: The idealized pixelated image. Since the image is radially symmetric, we've plotted a cut along the $x$ axis. All three fits from this image are almost perfect, as visible in the left panel and the blow-up in the right panel.}
\label{fig:perfectim}
\end{figure*}

Pixelation is one obvious difference between a real image and the idealistic image above. To demonstrate the effect of pixelation, we've generated a simple, pixelated image:
\begin{enumerate}
\item \pix
\item \flatilm
\item \simplepsf
\item \onesphere
\item \flatbkg
\item \nonoise
\end{enumerate}
Figure~\ref{fig:perfectim} shows such an image and the three algorithms' performances. The particle is centered exactly on the pixel at position ($x,y,z) = (10,10,10)$, it has a 5 pixel radius and is imaged with a Gaussian point-spread function with width $\sigma=1$. These parameters are typical for a 1 $\mu$m diameter sphere imaged with a confocal microscope at a 100x magnification, corresponding to a pixel size of 100 nm in the object space.

The heuristics look great! All of the methods fit accurately to $<$0.001 pixels, or 1 \angstrom for a typical 100 nm pixel size! What's the problem?

\begin{figure*}[ht]
\includegraphics[width=\textwidth]{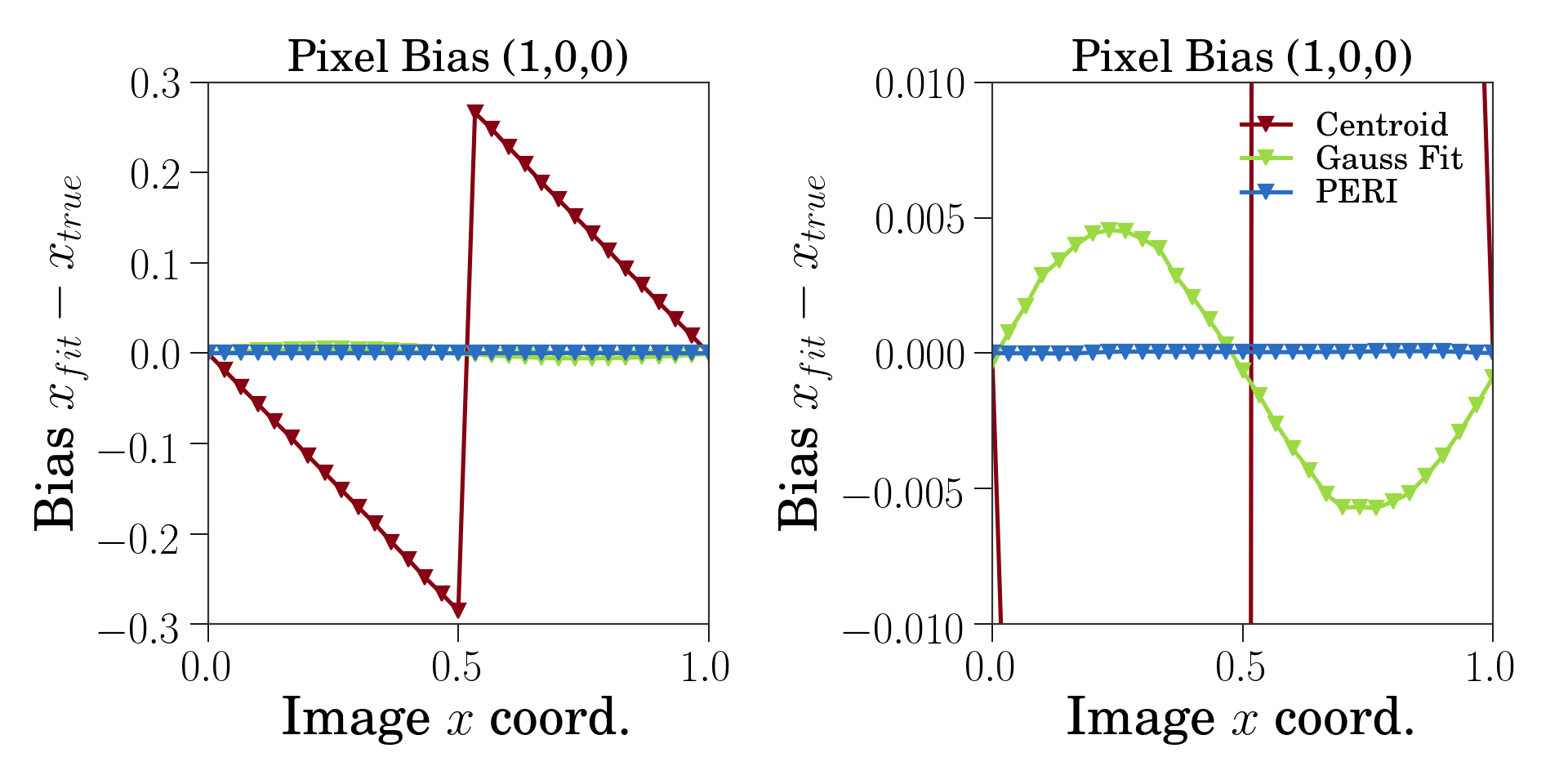}
\caption{Pixel bias, for particles shifted by a fraction of a pixel along the $(x,y,z)=(1,0,0)$ direction. The right panel is a zoomed in version of the left.}
\label{fig:pix001}
\end{figure*}

\begin{figure*}[ht]
\includegraphics[width=\textwidth]{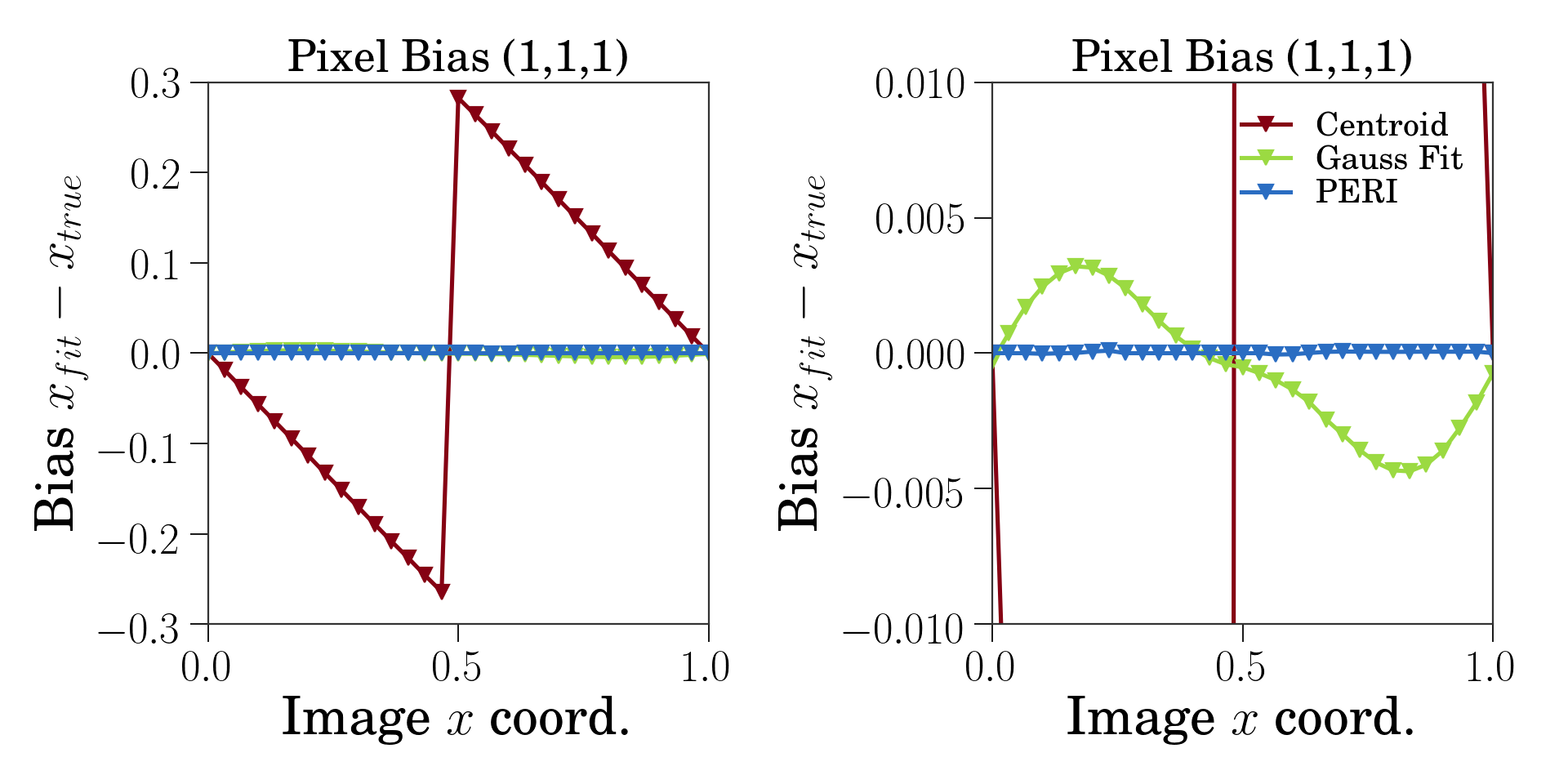}
\caption{Pixel bias, for particles shifted by a fraction of a pixel along the $(x,y,z)=(1,1,1)$ direction. The right panel is a zoomed in version of the left.}
\label{fig:pix111}
\end{figure*}

The problem starts to appear when we shift the particle by a fraction of a pixel, as demonstrated in figure~\ref{fig:pix001} and \ref{fig:pix111}. The measured positions are no longer correct but are biased from their correct values by a small amount. For this simple image, the bias only depends on the fractional displacement of the particle's position relative to one pixel, \text{i.e} the bias for a particle at $x=9.1$ and $x=10.1$ should be the same. In general, this pixel bias in each coordinate varies with position of all three coordinates in the unit cell of the pixel. To give an idea of what it looks like, we've plotted the bias in the $x$ position as a function of the particle's position along two different axes in the unit cell: the $(x,y,z)=(1,0,0)$ and $(1,1,1)$ axes, as shown in figures~\ref{fig:pix001} and \ref{fig:pix111}.

This problematic bias is known as pixel bias or pixel locking and is discussed extensively in the literature~\cite{Prasad1992, Cheezum2001, Feng2007, Berglund2008, Gao2009, Anthony2009, Besseling2015, Hearst2015, Burov2017}. The finite sampling from the pixelation breaks the translational invariance of the image and prevents simple heuristics from getting the exact answer. Due to the periodicity and mirror symmetry of the pixelated grid, the bias $\bm{b}(\bm{x})$ is both periodic and odd about the pixel center. Since the bias is odd, there is no bias for particles centered exactly on a pixel. The raw centroid method has significant pixel bias, as shown in figure~\ref{fig:pix001} and \ref{fig:pix111}. The Gaussian fitting does well, but it is not perfect. As visible in the figures, the Gaussian fitting method produces a bias in the fit of about 0.005 px, or 0.5 nm for a typical 100~nm pixel size. But these images have no noise, so the Cramer-Rao bound~\cite{Rao1945, Bierbaum2017} is 0, and a good method should get the exact answer. For these images, the imperfection is mostly intellectual, since 0.5 nm is a tiny error, but small errors like this can add up and more problems will appear with the Gaussian fitting method as we move on. Finally, in contrast to the two heuristic methods, the reconstruction method PERI locates the particle almost exactly, as it should. The tiny numerical errors of about $10^{-5}$ pixels are from slightly incomplete fit convergence, which is typical in most least-squares fitting algorithms.

\section{Single Particle: Varying Illumination}
\label{sec:ilm}
Real images have spatially-varying illumination, whether from dirt on the optics or from the light source itself. The gradient in illumination causes biases in the featuring if not accounted for, shifting the actual intensity centroid of the region and the best-fit Gaussian center. To illustrate this, we've generated and fit a series of images with pixelation and a simple illumination gradient:
\begin{enumerate}
\item \pix
\item \gentleilm
\item \simplepsf
\item \onesphereshort
\item \flatbkg
\item \nonoise
\end{enumerate}
To mitigate the effect of pixel bias, we've centered the particle exactly at $(x,y,z) = (10.0, 10.0, 10.0)$, on a point with zero pixel bias. Instead of being constant, the illumination has a slight gradient along the $x$ direction only. This gradient creates a bias in the measured positions along the $x$ direction, while leaving the measured positions along the $y$ and $z$ directions unbiased, due to symmetry.

\begin{figure*}[ht]
\includegraphics[width=\textwidth]{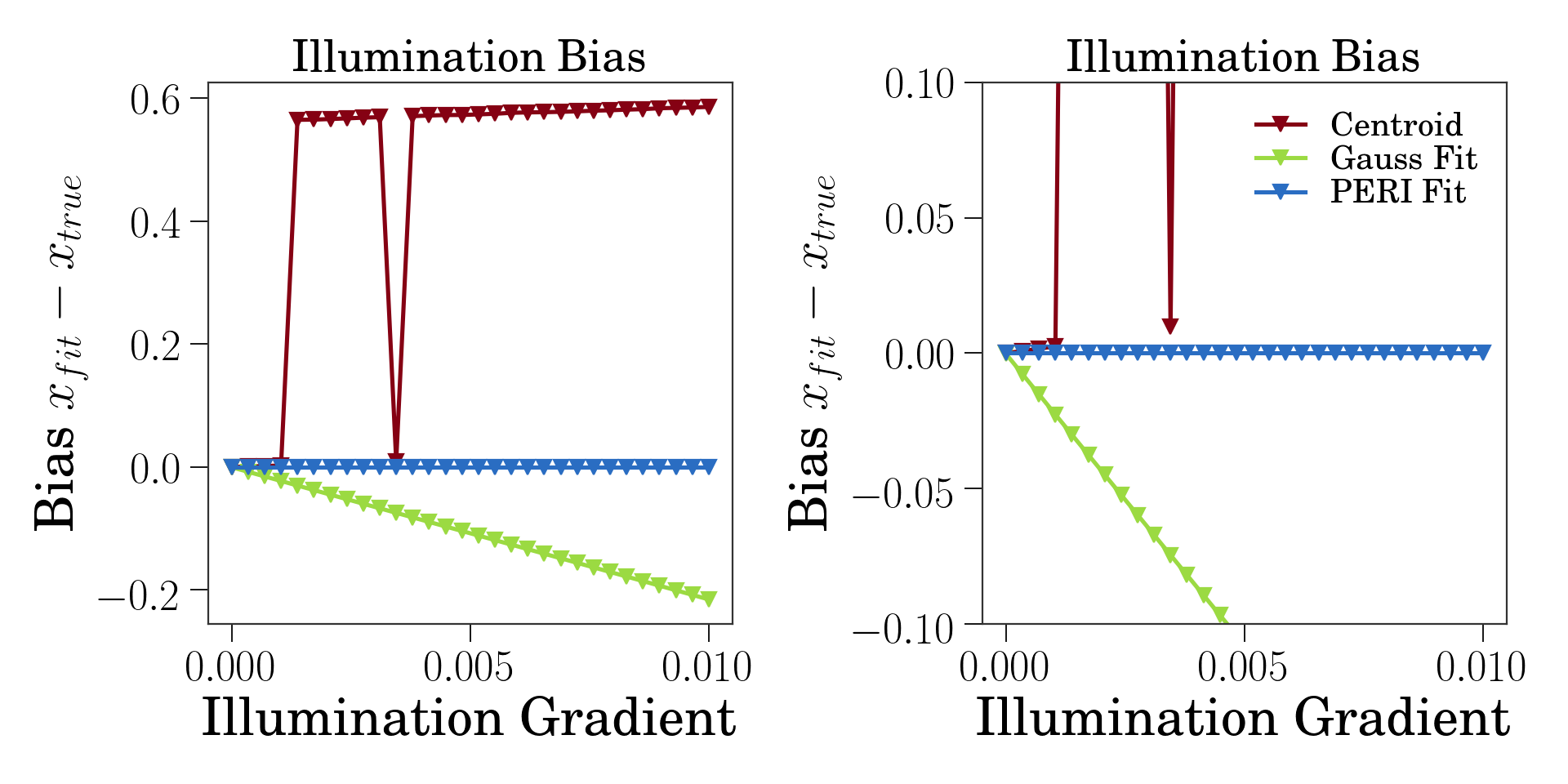}
\caption{Bias induced by a realistic illumination gradient. The ``Gauss-fit`` method produces a bias that is linear in the illumination gradient. While there is no bias for a gradient-free image, the bias increases to a sizable fraction of a pixel as the gradient increases. In contrast, \tp{} performs well sometimes and poorly other times. PERI properly fits the illumination and thus has no bias in the figure.
}
\label{fig:ilmbias}
\end{figure*}

These biases along the $x$-direction are illustrated in figure~\ref{fig:ilmbias}. Once again, PERI locates the particle perfectly, with tiny deviations of $\approx 10^{-9}$ which go to zero as the optimization continues to improve the reconstruction. In contrast, both heuristic methods produce a clear bias that worsens with increasing illumination gradient. The bias from fitting to a Gaussian increases linearly with the intensity gradient, as one might expect, whereas the bias from \tp{} bounces around discontinuously, seeming to do well at small illumination biases but extremely poorly at larger biases. At low illuminations \tp{} does well because it subtracts the background with a long-wavelength filtering. For the simple, linearly-varying illumination in these images, subtracting the background performs fairly well. The discontinuous jumps in bias may arise from a detail of the implementation: \tp{} functions by finding a region-of-interest about the particle and calculating the center of mass of that region. If the center of mass is different from the region's center by more than a pixel, then the region of interest is shifted and the center of mass calculated again. The process is iterated~\cite{Gao2009}, either until convergence or for a maximum number of times. However, the window shifts discontinuously, by a whole pixel at a time. (The discontinuous shift is present in both \tp{} v. 0.3.2 and in the Crocker-Grier-Weeks IDL code, when iteration is used~\footnote{In \tp{}, the iteration occurs in lines 345ff in trackpy/feature.py, in the function \texttt{\_refine}. In the IDL functions, iteration occurs in lines 388ff in feature.pro, in the function \texttt{feature}. There is no obvious any iteration in the 3D IDL tracking routine, \texttt{feature3d}.}). This discontinuous shift could be what causes the discontinuous jumps in error~\cite{Berglund2008}.

\begin{figure*}[ht]
\includegraphics[width=0.667\textwidth]{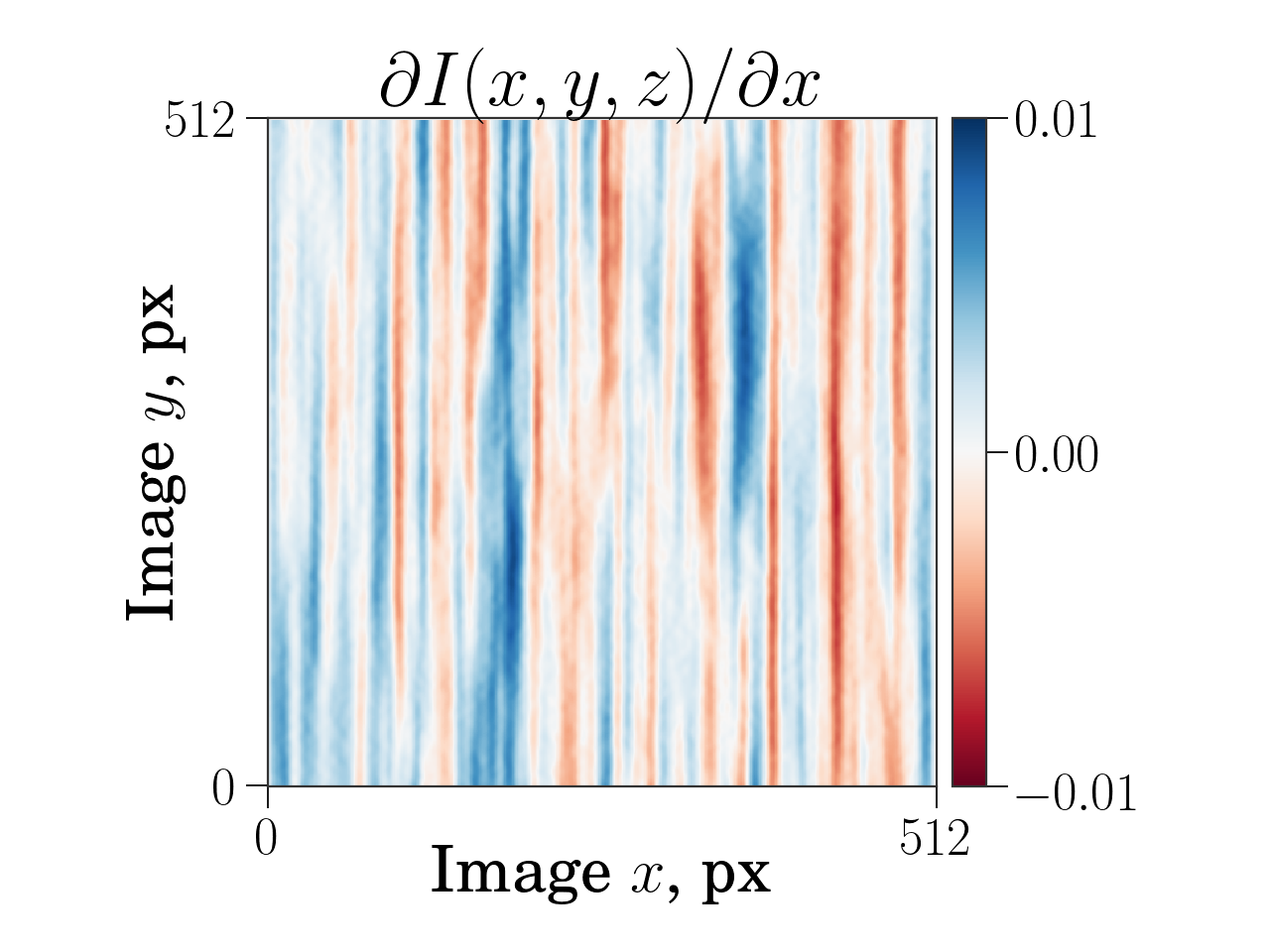}
\caption{
Measured illumination gradients along the $x$-direction, from an image of fluorescent dye without any particles in it; the illumination is normalized to have a mean of 1. The gradient field shown here has been smoothed twice: once by the blurring action of the point-spread function in the image acquisition, and again with a Gaussian filter to remove some of the noise. The illumination gradients in this image after smoothing on the scale of the particle size are about 0.01 / px, the scale in figure~\ref{fig:ilmbias}.
}
\label{fig:ilmdata}
\end{figure*}

Why not just subtract the spatially-varying illumination? This seems like an obvious solution, especially given the performance of \tp{} at low illumination gradients. For a simple image with a slowly-varying illumination (like the one we've generated), subtracting the background will indeed work. However, for realistic images the illumination can vary rapidly enough where subtracting it is not feasible, even varying on the scale of the particle size. As an example, figure~\ref{fig:ilmdata} shows the illumination gradient measured from an experimentally-measured illumination field \textit{after} smoothing on the scale of a particle's size. Even after this smoothing, the illumination gradient still varies considerably. For rapidly-varying illumination gradients like that shown in the figure, removing the long-wavelength variations in intensity will still miss all the short wavelength structure. Even worse, in this image the illumination even changes on the scale of the particle size, and it's not even clear what removing this rapidly-varying illumination would entail.

\section{Single Particle: Asymmetric Point-Spread Function}
\label{sec:psf}
So far all the images have been generated with a simple Gaussian point-spread function. However, real confocal point-spread functions are not Gaussians. Confocal point-spread functions have long tails. Confocal point-spread functions are anisotropic, with longer tails along the optical axis ($z$) than the other two. Most importantly, confocal point-spread functions are aberrated under realistic imaging conditions. Whenever the index of refraction of the sample is mismatched from the optical train's design, then there will be aberrations present in the imaging~\cite{Hell1993, Gibson1991, Juvskaitis1998, Booth1998, Diaspro2001, Jenkins2008, Nasse2010, Lin2014}. This aberration
\begin{itemize}
\item shifts the center of the point-spread function,
\item darkens the portions of the image that are deeper in the sample,
\item broadens the point-spread function, and
\item skews the point-spread function, making it asymmetric between $+z$ and $-z$.
\end{itemize}
All of these effects can cause biases in measured data. The shifting of the point-spread function's center effectively shrinks the optical $z$ direction compared to the in-plane directions, which will bias the particle $z$ positions~\cite{Lin2014, Bierbaum2017}. The darkening of deeper regions in the image will induce biases in the same manner as a varying illumination. And the broadening of the point-spread function increases the biases induced by nearby particles, as detailed in the next section.

\begin{figure*}[ht]
\includegraphics[width=\textwidth]{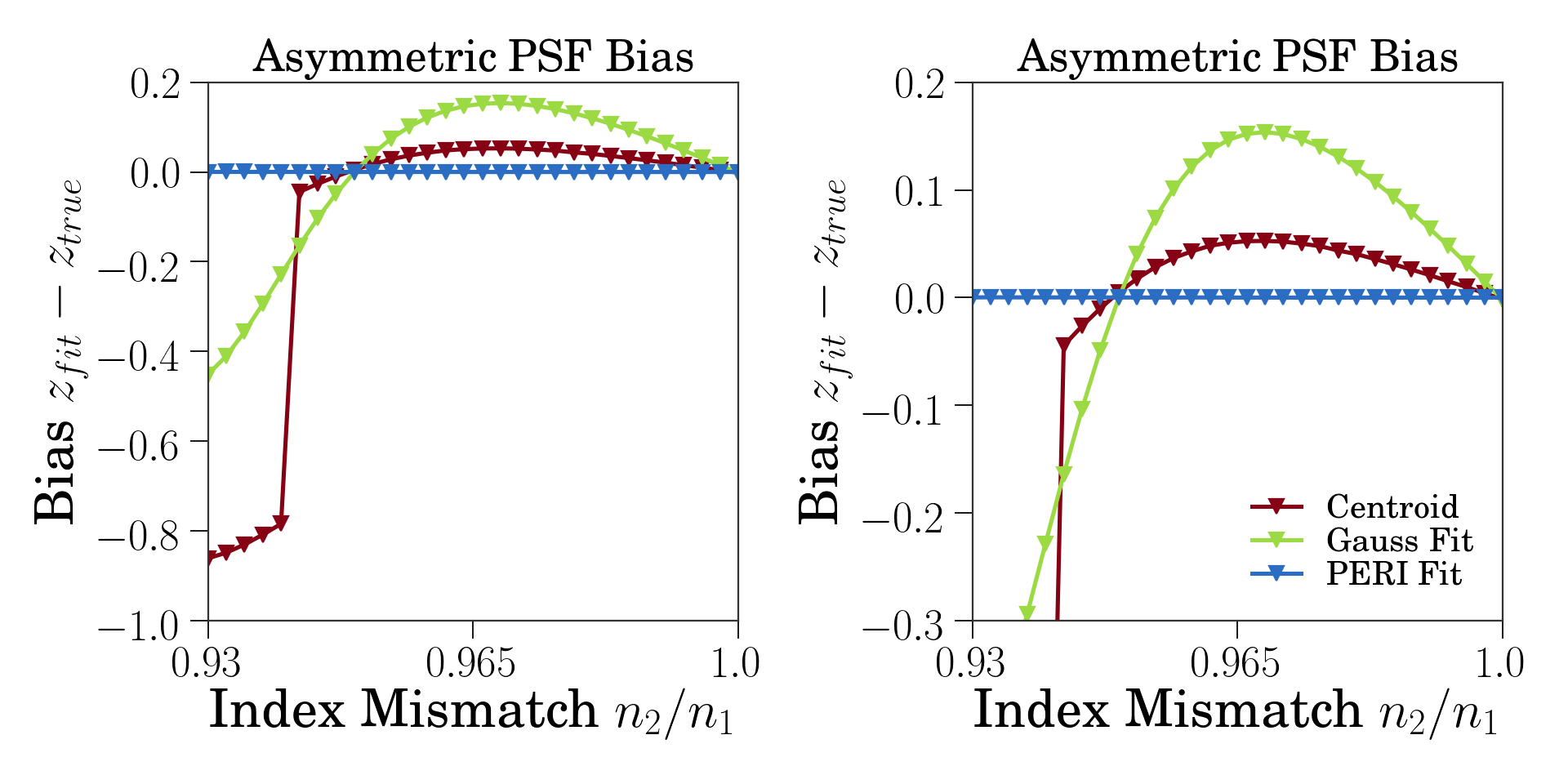}
\caption{
Bias induced by a real point-spread function. As the aberration increases, so does the bias in both heuristic methods, in a complex, non-monotonic manner. In contrast, PERI properly fits the point-spread function and thus has little to no bias in the figure. An index-matched sample of silica particles ($n_2 \approx 1.44$) imaged on a typical microscope (design index $n_1 \approx 1.52$) sits at $n_2/n_1 \approx 0.95$; an index-matched sample of PMMA particles ($n_1 \approx 1.49$) sits at $n_2/n_1 \approx 0.98$. The right panel is a zoomed in version of the left.
}
\label{fig:psfbias}
\end{figure*}

But even just the asymmetry of the point-spread function can significantly bias the extracted positions and radii. To demonstrate this one effect, we've generated images with:
\begin{enumerate}
\item \pix
\item \flatilm
\item \realpsf
\item \onesphereshort
\item \flatbkg
\item \nonoise
\end{enumerate}
Positioning the sphere exactly at a pixel center and illuminating it with uniform illumination prevents any pixel bias or illumination gradient bias, leaving the only source of bias as the point-spread function. To isolate the effects of the point-spread function's asymmetry, we've generated images with the deeper regions re-normalized to the same brightness as the shallower regions, and with the compression of the optical axis removed \footnote{Incidentally, the current implementation of PERI generates its images in a similar manner. We fold the focal shift into a global $z$-scale and the varying intensity into a combination of the illumination and background. This makes the fit landscape simpler, decoupling the strong variation of the cost with illumination from the weaker variation with the point-spread function.}.

We generate images using the point-spread function of a line scanning confocal microscope  under realistic imaging conditions. We calculate the point-spread function for a microscope with a lens numerical aperture of 1.4, with illumination light of 488 nm wavelength and detected fluorescence at 550 nm wavelength, the nominal values of our current imaging setup. The pixel size in object space is 100 nm, which is typical for 100x magnification. The lens is perfect aside from aberrations induced by the index mismatch between the sample and the optics. These aberrations increase with both the index mismatch of the sample and the distance from the optical interface~\cite{Hell1993}. Here, we control the aberrations by varying the index mismatch $n_2/n_1$ of the sample ($n_2$) to the optics ($n_1$) while keeping the particle's center at a fixed $z$ position 1~$\mu$m above the optical interface. In the limit $n_2/n_1=1$, there is no aberration and the PSF is symmetric with $z\rightarrow -z$. As $n_2/n_1$ decreases away from $1$, the aberration increases and the point-spread function becomes increasingly asymmetric. Typical images of colloidal suspensions are taken with $n_2/n_1 \approx 0.95 - 0.98$, corresponding to index-matched silica or PMMA.

Figure~\ref{fig:psfbias} shows the extracted particle locations from \tp{}, from fitting to a Gaussian, and from PERI. When $n_2/n_1 = 1$, the symmetry of the PSF ensures that both centroid methods and fitting to a Gaussian perform well. As $n_2/n_1$ decreases away from 1, the PSF becomes increasingly aberrated and the heuristic methods perform considerably worse, with biases up to 0.8 px (10-80 nm). In contrast, PERI always accurately measures the single particle's position. At very large aberrations ($n_2/n_1 < 0.95$), PERI performs relatively poorly because the fit landscape of the point-spread function becomes tortuous, with some possible local minima. However even in these highly-aberrated images, PERI locates particles to about $10^{-4}$ pixels. As the image becomes less aberrated ($n_2/n_1 > 0.95$), the simpler fit landscape makes it easier for PERI to properly fit the image. Here, PERI locates particles to about $10^{-5}$ pixels, despite the difficulty of fitting both the PSF (7 parameters) and the particle's position and radius (4 parameters) with only one particle in the image.

\section{Adjacent Particles}
So far we have only examined biases in images of isolated particles. But frequently as researchers we're interested in experiments with many particles. When particles are close, their diffractive blurs from the point-spread function can overlap. This overlap interferes with heuristic measures, biasing the extracted positions to be fit closer or farther together than they actually are~\cite{Baumgartl2005, Polin2007}.

To demonstrate this, we've generated some data of two particles with surface-to-surface separations of an amount $\delta$ in the $x$-direction.
\begin{enumerate}
\item \pix
\item \flatilm
\item \realpsf
\item \twosphere
\item \flatbkg
\item \nonoise
\end{enumerate}
Since the amount of bias due to an adjacent particle depends on the details of the point-spread function, we've used a physical point-spread function with an index mismatch of $n_2/n_1=0.95$ (typical for silica particles in an index-matched solution). While the PSF is asymmetric in $z$ it is still symmetric with $x \rightarrow -x$ and $y \rightarrow -y$, leaving the image of an isolated particle un-biased in $x$. Likewise, the uniform illumination prevents any illumination gradient bias. However, since we've generated data at separations of a fraction of a pixel, the heuristic methods have some pixel bias present in them.

\begin{figure*}[ht]
\includegraphics[width=\textwidth]{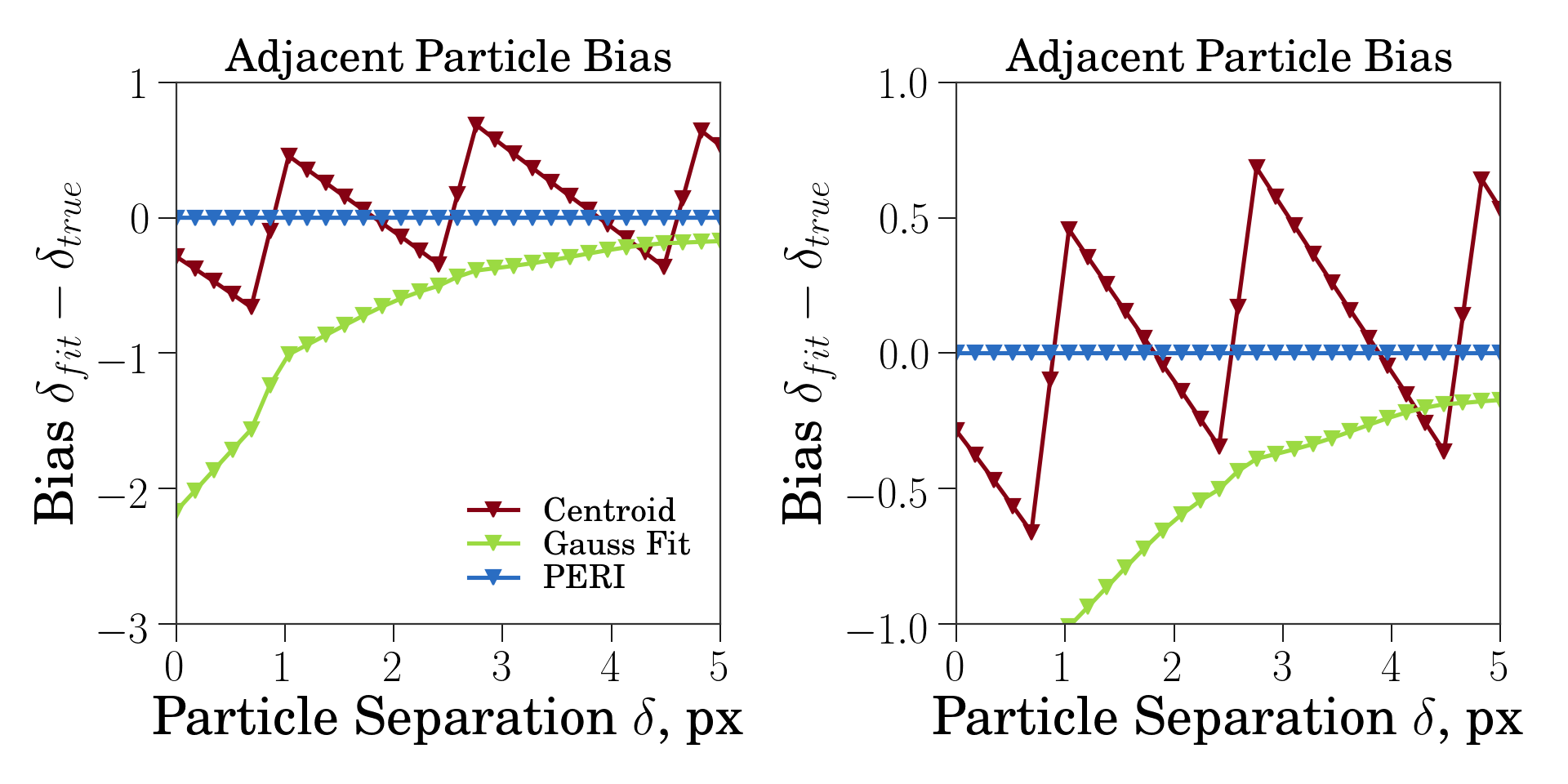}
\caption{
Position bias induced by a second nearby particle. Both centroid methods and fitting to a Gaussian have biases that are a significant fraction of a pixel or greater, even out to separations of 5 pixels (500 nm). In contrast, PERI fits the particle positions exactly (errors of a few times $10^{-4}$ pixels, due to fit convergence). The right panel is a zoom in of the left.
}
\label{fig:sepbias}
\end{figure*}

The heuristic methods perform quite poorly when there is more than one particle, as shown in figure~\ref{fig:sepbias}. When the particles are in contact (separation of $0$ pixels), both the heuristic methods have serious errors. Fitting a Gaussian separately to each particle returns sizable errors -- about 2 px near contact! The errors from \tp{} are nearly as bad, fluctuating with a magnitude of about 0.5 px. Even at a large separation of 5 px (\textit{i.e.} one particle radius), the biases are still considerable -- 0.2 px or 20 nm for fitting and still about 0.5 px for \tp{}. In contrast, PERI identifies the particle positions accurately, with minuscule errors of a few times $10^{-5}$ pixels (corresponding to 1 pm) due to the fit converging.

\section{Realistic Image}
We've examined several realistic sources of bias separately. What happens when we include these all together, in a realistic image?

To examine this, we've generated two highly realistic images, one with noise and one without:
\begin{enumerate}
\item \pix
\item \realilm
\item \realpsf
\item Many (123) spheres randomly distributed with varying radii, with some of the particles partially outside the image.
\item Spatially-varying background.
\item No noise (first) then with noise (second).  \end{enumerate}
To keep these images as realistic as possible, we first used PERI to fit a real image of a moderately-dilute suspension (volume fraction $\phi \approx 0.13$) of colloidal spheres. We then used the best-fit image as the generated data, cropped to a smaller size for speed and convenience. For the noisy image, we added white Gaussian noise at a signal-to-noise ratio of 30.

\begin{figure*}
    \includegraphics[width=\textwidth]{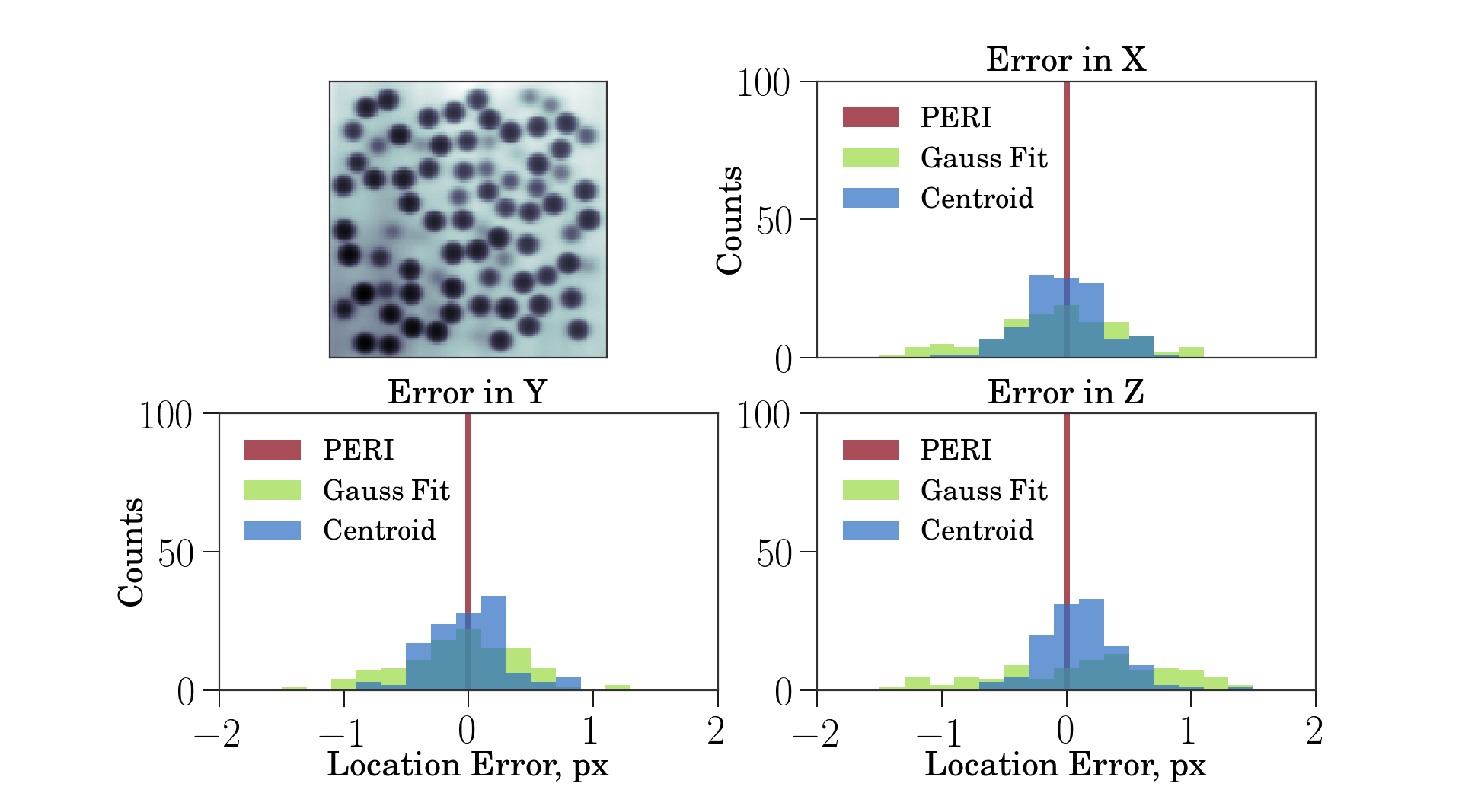}
    \caption{
    Position bias for a realistically generated image without noise (cross-section shown in upper-left panel). Even without noise, effects such as a spatially-varying illumination, nearby particles, pixel bias, and the complex point-spread function cause significant biases in the heuristic methods. In contrast, PERI accurately measures the $x$, $y$, and $z$ positions.
    }
    \label{fig:realim}
\end{figure*}

\begin{figure*}
    \includegraphics[width=\textwidth]{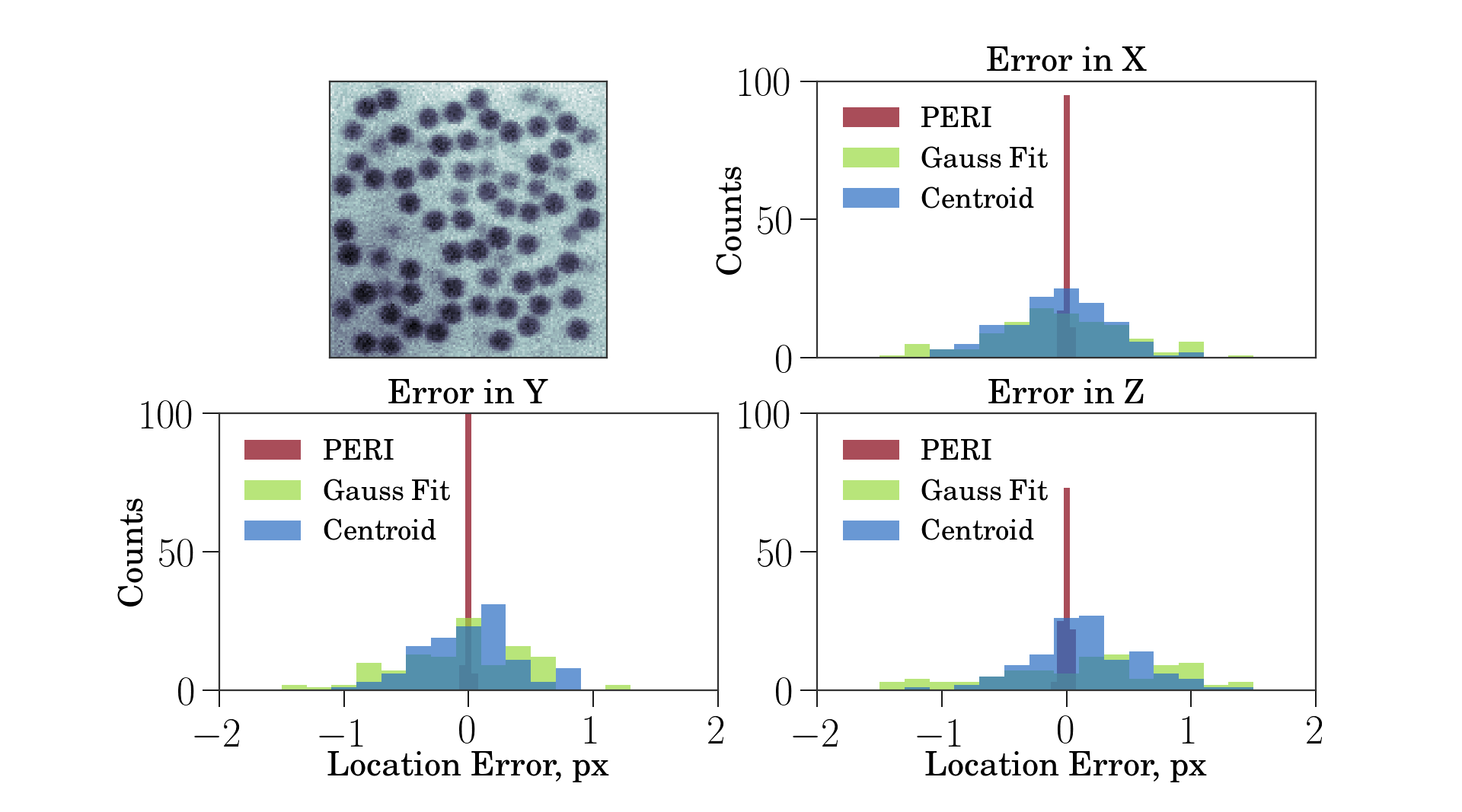}
    \caption{
    Position bias for a realistically generated image with noise (cross-section shown in upper-left panel). Since most of the problem with the heuristic methods is due to systematic biases, adding noise does not greatly worsen their results. Interestingly, the errors in the heuristic featuring methods differ considerably from the noise-free dataset, by more than would be expected from just the addition of noise. The noise worsens the accuracy of PERI to the 1-nm scale of the Cramer-Rao bound for this image.
    }
    \label{fig:noiseim}
\end{figure*}

By using a generated image rather than a real image, we can know the true positions and radii of the particles and we can directly measure biases in the algorithms on the particle-scale. In addition, by generating the same image with and without noise, we can test the relative effects of random noise versus systematic biases on the three localization algorithms. Since the fitted image accounts for all but $10^{-5}$ of the signal in the image, we can be certain that our generated image is highly realistic, and that the biases observed while analyzing our generated image are representative of biases that would be observed while analyzing real data. We measure the biases by analyzing the generated images with each of the three algorithms, as before, and looking at the differences between the true positions and the measured positions. However, in practice most researchers are interested in relative separations or displacements of particles rather than absolute positions. As such, we've subtracted off an overall shift in $(x, y, z$) for the entire image, such as would arise from a translation of the coordinate system, before calculating errors~\footnote{Removing the shift mostly affects the particle $z$ positions. As discussed in section~\ref{sec:psf}, spherical aberration creates an asymmetry in the point-spread function. Increasing this asymmetry looks very similar to shifting the particles in $z$, which creates additional problems for all the heuristic methods and a slow mode in the fit for PERI.}. This overall shift is listed in tables~\ref{tbl:noisefree} and \ref{tbl:noise}, but the shift is removed for the error columns in the tables and for the histograms in figures~\ref{fig:realim}-\ref{fig:noiseim}.

Once again, the heuristic methods are limited, despite the absence of noise in the image. The centroid method \tp{} has considerable errors of $(0.32, 0.32, 0.36)$ px or $(32, 32, 36)$ nm in ($x$, $y$, $z$). Fitting to a Gaussian performs no better, returning errors in ($x$, $y$, $z$) of $(0.7, 0.56, 1.3)$ px or $(70, 56, 130)$ nm. In contrast, measuring particle positions via a detailed reconstruction of the image accounts for the confounding sources of bias. The reconstruction method PERI accurately measures the particle positions, with an ($x$, $y$, $z$) error of (4e-4, 2e-4, 3e-4) px or $(0.04, 0.02, 0.03$) nm, with the slight imperfection arising from a soft direction in the fit landscape causing slow convergence~\cite{Transtrum2010}. Running the fit for additional time slightly improves the image reconstruction and the accuracy in the extracted parameters.

Adding noise at a signal-to-noise ratio of $30$ affects these results only slightly, as shown in figure~\ref{fig:noiseim}. The errors in the heuristic methods increase only moderately, as most of their errors arise from systematic sources and not from noise. And fitting the data to a generative model performs well in the presence of noise, localizing particles in ($x$, $y$, $z$) with $(0.02, 0.02, 0.03)$ px errors, at the Cramer-Rao lower bound.

\begin{table}[h]
    \begin{tabular}{r||c|c|c|c|c|c}
        Method & X-Shift & X-Error & Y-Shift & Y-Error & Z-Shift & Z-Error \\
        \hline \hline
        Centroid & -0.016 & 0.32 & 0.0013 & 0.32 & 0.14 & 0.36 \\ \hline
        Gauss-fit & -0.19 & 0.7 & -0.086 & 0.56 & -0.33 & 1.3 \\ \hline
        PERI & -9.8e-05 & 0.00039 & -6.1e-05 & 0.00018 & -0.00028 & 0.0024 \\
    \end{tabular}
    \caption{\textbf{Biases in realistic, noise-free images.} Biases for all
        three methods on the realistic images along x, y, and z, with the bias
        separated into an overall shift and the fluctuations about that shift.}
    \label{tbl:noisefree}
\end{table}

\begin{table}[h]
    \begin{tabular}{r||c|c|c|c|c|c}
        Method & X-Shift & X-Error & Y-Shift & Y-Error & Z-Shift & Z-Error \\
        \hline \hline
        Centroid & -0.063 & 0.41 & 0.008 & 0.37 & 0.16 & 0.48 \\ \hline
        Gauss-fit & -0.16 & 0.69 & -0.093 & 0.57 & -0.37 & 1.3 \\ \hline
        PERI & -0.0029 & 0.021 & 0.0018 & 0.017 & -0.0034 & 0.03 \\
    \end{tabular}
    \caption{\textbf{Biases in realistic, noisy images.} Biases for all three methods on the realistic images along
        x, y, and z, with the bias separated into an overall shift and
        the fluctuations about that shift.}
    \label{tbl:noise}
\end{table}

\section{Closing Thoughts}

\subsection{Why does a generative model approach work?}
Why does the reconstruction-based approach of PERI accurately measure particle positions while heuristic methods perform so poorly in comparison? As we've alluded to throughout this document, PERI performs well because it includes more of the possible systematic effects.

But there is a deeper reason why PERI includes more of these possible systematic effects: we fit a generative model, and \textit{we compare the fit to the data}. By looking carefully at the difference between the fit and the data, we can understand whether or not our model is complete. If it's not complete, usually the fit residuals or the extracted parameters themselves provide a clue as to what is missing. For instance, at early stages in the development we thought that a relatively low-order polynomial was sufficient to describe our spatially-varying illumination. However, our fit residuals oscillated with stripes in the $x$-direction, and the particle radii oscillated in phase with the stripes. Switching to a higher-order stripe-like illumination smoothed out the residuals and removed the stripe-like radii bias. Likewise, we initially thought that a Gaussian approximation to the point-spread function would suffice to measure particle radii~\cite{Zhang2007}. However, there were systematic variations in the particle radii along the optical axis ($z$), and the residuals contained strong rings around the particles. Allowing the width of the Gaussian to vary with $z$ provided a partial remedy, but we still found significant radii biases and still saw rings around the particles in the residuals. Switching to an exact model of a physical point-spread function both removed the rings in the fit residuals and significantly improved our radii measurements. By reconstructing the image, we were able to see which physical effects were missing from our model and possibly causing systematic errors. This process of model selection and model updating is discussed in detail in the supplemental information of ref.~\cite{Bierbaum2017}. Once the source of the systematic error is known, it is conceptually straightforward to implement the additional physics in the generative model.

Moreover, with a generative-model framework, it is easy to test whether additional physics that is too difficult to include in the model introduces biases -- simply generate a realistic image with the additional effects and analyze it with the generative model. For instance, we PERI developers were worried that Brownian motion might bias the extracted parameters, since during the image exposure time particles diffuse by much more than the Cramer-Rao bound for our images. As a random trajectory, Brownian motion is too complicated to attempt to fit in a generative model. However, it is comparatively easy to generate an image of a diffusing particle. From the generated data, we were able to discern that the bias due to Brownian motion is extremely weak, many times smaller than the noise floor from the Cramer-Rao bound for typical confocal images~\cite{Bierbaum2017}.

In contrast, heuristic approaches provide no clear way to completely incorporate sources of systematic errors or even to know at what level they matter. How can one guess from a list of positions that the uneven illumination causes a slight bias? That even distant particles slightly bias a particle's position due to the long-tails of the point-spread function? Even if something about the extracted positions or prior knowledge of the image formation tips off a researcher that there are systematic errors in their data, how can a heuristic method accurately account for them? For simple sources of bias, such as uneven illumination, occasionally adding another component to the heuristic such as fitting locally to a polynomial~\cite{Rogers2007} can improve the extracted positions. But correctly accounting for even these simple sources of bias is not straightforward -- \textit{e.g.} what order polynomial should be used? Over what region of interest should it be fit? Worse, it becomes increasingly difficult to compensate a heuristic algorithm for complex sources of bias, such as aberration due to the point-spread function. In contrast, a generative model allows a directed approach to incorporating sources of systematic errors -- based on physical knowledge of the image formation -- with residuals in the fit announcing any additional sources of bias.

In light of this, the excellent performance of a reconstruction over a heuristic arises mostly from practical reasons. In principle, it could be possible to dream up a heuristic method which accurately accounts for all conceivable sources of bias, or to discover a post-processing method which removes all those possible sources of bias. However, it is near impossible to conceive of a heuristic method which will be bias-free for all images, and there is no rigorous way to check the quality of the estimated positions against the raw image data. Conversely, while any generative model for image formation will not include every physical effect, the quality of the model can be checked by comparing the difference between the reconstruction and the raw image data. If the difference is not Gaussian white noise, then the model and its measurements are suspect, and the model and its reconstruction must be improved. If the difference is indistinguishable from Gaussian white noise or nearly so, then the model and its measurements should represent reality reasonably well. The generative model of PERI captures most, but not quite all, of the signal in real microscope images. This slight incompleteness in PERI's model results in nm-scale biases in the extracted positions and radii from real images~\cite{Bierbaum2017}.

\subsection{Relative Displacements vs. Absolute Positions}

While a heuristic method will not measure particle positions or particle separations with nanometer accuracy, that doesn't necessarily mean that a heuristic method can't accurately measure relative \textit{displacements} of individual particles over time. As we've shown above, most of the sources of error in measuring a particle's position (or radius) arise from systematic difference in the image formation. In general, these sources of systematic bias will depend on the configuration of particles and the properties of the optics. For most ``reasonable'' algorithms, changing the sources of systematic errors will continuously and smoothly change the reported particle locations and properties, \textit{i.e.} the bias $\bm{b}(\bm{x})$ is a smooth function of the configuration of particles $\bm{x}$. Thus, two consecutive images with only small changes in particle configuration between images will have almost identical systematic errors, which will cancel when the displacement is calculated~\footnote{A counter-example would be the large discontinuities in the biases measured by \tp{} shown in the figures.}. In other words, if the true position is $\bm{x_t}(t)$ at time $t$, and the measured position $\bm{x_m}(t)$ is the true position plus a bias $\bm{b}(\bm{x_t}(t))$, then the measured relative displacement $\Delta \bm{x_m} \equiv \bm{x_m}(t_1) - \bm{x_m}(t_0)$ between the positions at two separate times $t_1$ and $t_0$ is 
\begin{align*}
\Delta \bm{x_m} & = [\bm{x_t}(t_1) + \bm{b}(\bm{x_t}(t_1))] - [\bm{x_t}(t_0) + \bm{b}(\bm{x_t}(t_0))] \\
& = \Delta \bm{x_t} + \left[ \bm{b}(\bm{x_t}(t_1)) - \bm{b}(\bm{x_t}(t_0)) \right] \;,
\end{align*}
where $\Delta \bm{x_t}$ is the true displacement. If the two configurations $\bm{x_t}(t_1)$ and $\bm{x_t}(t_2)$ are close, then $\bm{b}(\bm{x_t}(t_1)) \approx \bm{b}(\bm{x_t}(t_0))$, and the measured displacement will be approximately correct.

As an obvious corollary, test images with small relative displacements of particles or test images of particles fixed in place are not good measures of errors in particle localization. Since real errors in measured particle positions mostly arise from systematic sources of error, changing the configuration of particles by a small amount doesn't significantly change any sources of systematic errors. As a result, using test images with little or no configurational changes of particles to estimate localization errors -- such as repeatedly imaging particles fixed in place -- will severely underestimate the scale of systematic errors.

\section{Conclusion}

At this point, we hope you're convinced that heuristic methods give approximate results. But that doesn't mean that heuristic results aren't useful. Fitting a complete generative model to data takes many, many times longer than running a simple heuristic like centroid identification. If you don't need the accuracy, why waste the time? Even if you \textit{do} need the accuracy, it might be worthwhile to analyze your data first with a fast heuristic method. There is no point in wasting days to weeks of computer time only to realize that your sample was bad or the time resolution of your data was too slow. Analyzing the data with a fast heuristic method first can help you avoid missteps like these. However, if you want or need highly-accurate positions from your data, then a reconstructive generative model approach when correctly used will provide them.

\acknowledgments
We would like to thank C. Clement, A. Alemi, N. Lin, L. Bartell, M. Ramaswamy, and P. McEuen for useful discussions. This work was supported in part by NSF DMR-1507607 and ACS PRF 56046-ND7.

\bibliographystyle{plain}


\begin{thebibliography}{10}

\bibitem{trackpy}
Daniel~B. Allan, Thomas~A. Caswell, and Nathan~C. Keim.
\newblock Trackpy v0.2, May 2014.

\bibitem{Andersson2008}
Sean Andersson.
\newblock Localization of a fluorescent source without numerical fitting.
\newblock {\em Optics express}, 16(23):18714--18724, 2008.

\bibitem{Anthony2009}
S.~M. Anthony and S.~Granick.
\newblock Image analysis with rapid and accurate two-dimensional gaussian
  fitting.
\newblock {\em Langmuir}, 25:8152--8160, 2009.

\bibitem{Baumgartl2005}
J{\"o}rg Baumgartl and Clemens Bechinger.
\newblock On the limits of digital video microscopy.
\newblock {\em EPL (Europhysics Letters)}, 71(3):487, 2005.

\bibitem{Berglund2008}
Andrew~J Berglund, Matthew~D McMahon, Jabez~J McClelland, and J~Alexander
  Liddle.
\newblock Fast, bias-free algorithm for tracking single particles with variable
  size and shape.
\newblock {\em Optics express}, 16(18):14064--14075, 2008.

\bibitem{Besseling2015}
T.~H. Besseling, M.~Hermes, A.~Kuijk, B.~de~Nijs, T-S. Deng, M.~Dijkstra,
  A.~Imhof, and A.~van Blaaderen.
\newblock Determination of the positions and orientations of concentrated
  rod-like colloids from 3d microscopy data.
\newblock {\em J. Phys. Condens. Matter}, 27:194109, 2015.

\bibitem{Bierbaum2017}
Matthew Bierbaum, Brian~D Leahy, Alexander~A Alemi, Itai Cohen, and James~P
  Sethna.
\newblock Light microscopy at maximal precision.
\newblock {\em Physical Review X}, 7(4):041007, 2017.

\bibitem{Booth1998}
Martin~J Booth, MAA Neil, and T~Wilson.
\newblock Aberration correction for confocal imaging in
  refractive-index-mismatched media.
\newblock {\em Journal of Microscopy}, 192(2):90--98, 1998.

\bibitem{Burov2017}
Stanislav Burov, Patrick Figliozzi, Binhua Lin, Stuart~A Rice, Norbert~F
  Scherer, and Aaron~R Dinner.
\newblock Single-pixel interior filling function approach for detecting and
  correcting errors in particle tracking.
\newblock {\em Proceedings of the National Academy of Sciences},
  114(2):221--226, 2017.

\bibitem{Cheezum2001}
Michael~K Cheezum, William~F Walker, and William~H Guilford.
\newblock Quantitative comparison of algorithms for tracking single fluorescent
  particles.
\newblock {\em Biophysical journal}, 81(4):2378--2388, 2001.

\bibitem{Chenouard2014}
Nicolas Chenouard, Ihor Smal, Fabrice De~Chaumont, Martin Ma{\v{s}}ka, Ivo~F
  Sbalzarini, Yuanhao Gong, Janick Cardinale, Craig Carthel, Stefano Coraluppi,
  Mark Winter, et~al.
\newblock Objective comparison of particle tracking methods.
\newblock {\em Nature methods}, 11(3):281--289, 2014.

\bibitem{Crocker1996}
J.~C. Crocker and D.~G. Grier.
\newblock Methods of digital video microscopy for colloidal studies.
\newblock {\em J. Colloid and Interface Science}, 179:298, 1995.

\bibitem{Diaspro2001}
A.~Diaspro.
\newblock {\em Confocal and two-photon microscopy: Foundations, applications,
  and advances}.
\newblock Wiley-Liss, Hoboken, New Jersey, 2001.

\bibitem{Feng2007}
Yan Feng, J~Goree, and Bin Liu.
\newblock Accurate particle position measurement from images.
\newblock {\em Review of scientific instruments}, 78(5):053704, 2007.

\bibitem{Gao2009}
Yongxiang Gao and Maria~L Kilfoil.
\newblock Accurate detection and complete tracking of large populations of
  features in three dimensions.
\newblock {\em Optics express}, 17(6):4685--4704, 2009.

\bibitem{Gibson1991}
Sarah~Frisken Gibson and Frederick Lanni.
\newblock Experimental test of an analytical model of aberration in an
  oil-immersion objective lens used in three-dimensional light microscopy.
\newblock {\em JOSA A}, 8(10):1601--1613, 1991.

\bibitem{Hearst2015}
R~Jason Hearst and Bharathram Ganapathisubramani.
\newblock Quantification and adjustment of pixel-locking in particle image
  velocimetry.
\newblock {\em Experiments in Fluids}, 56(10):191, 2015.

\bibitem{Hell1993}
S~Hell, G~Reiner, C~Cremer, and Ernst~HK Stelzer.
\newblock Aberrations in confocal fluorescence microscopy induced by mismatches
  in refractive index.
\newblock {\em Journal of microscopy}, 169(3):391--405, 1993.

\bibitem{Jenkins2008}
Matthew~C Jenkins and Stefan~U Egelhaaf.
\newblock Confocal microscopy of colloidal particles: towards reliable, optimum
  coordinates.
\newblock {\em Advances in colloid and interface science}, 136(1):65--92, 2008.

\bibitem{Jensen2016}
Katharine~E Jensen and Nobutomo Nakamura.
\newblock Note: An iterative algorithm to improve colloidal particle locating.
\newblock {\em Review of Scientific Instruments}, 87(6):066103, 2016.

\bibitem{Juvskaitis1998}
R~Ju{\v{s}}kaitis and T~Wilson.
\newblock The measurement of the amplitude point spread function of microscope
  objective lenses.
\newblock {\em J. Microsc}, 189(1):8--11, 1998.

\bibitem{Leocmach2013}
Mathieu Leocmach and Hajime Tanaka.
\newblock A novel particle tracking method with individual particle size
  measurement and its application to ordering in glassy hard sphere colloids.
\newblock {\em Soft Matter}, 9(5):1447--1457, 2013.

\bibitem{Lin2014}
Neil~YC Lin, Jonathan~H McCoy, Xiang Cheng, Brian Leahy, Jacob~N Israelachvili,
  and Itai Cohen.
\newblock A multi-axis confocal rheoscope for studying shear flow of structured
  fluids.
\newblock {\em Review of Scientific Instruments}, 85(3):033905, 2014.

\bibitem{Lu2013}
Peter~J Lu, Maor Shutman, Eli Sloutskin, and Alexander~V Butenko.
\newblock Locating particles accurately in microscope images requires
  image-processing kernels to be rotationally symmetric.
\newblock {\em Optics express}, 21(25):30755--30763, 2013.

\bibitem{Nasse2010}
M.~J. Nasse and J.~C. Woehl.
\newblock Realistic modeling of the illumination point spread function in
  confocal scanning optical microscopy.
\newblock {\em J. Opt. Soc. Am. A}, 27:295--302, 2010.

\bibitem{Parthasarathy2012}
R.~Parthasarathy.
\newblock Rapid, accurate particle tracking by calculation of radial symmetry
  centers.
\newblock {\em Nat. Methods}, 9:724--726, 2012.

\bibitem{Polin2007}
Marco Polin, David~G Grier, and Yilong Han.
\newblock Colloidal electrostatic interactions near a conducting surface.
\newblock {\em Physical Review E}, 76(4):041406, 2007.

\bibitem{Prasad1992}
AK~Prasad, RJ~Adrian, CC~Landreth, and PW~Offutt.
\newblock Effect of resolution on the speed and accuracy of particle image
  velocimetry interrogation.
\newblock {\em Experiments in Fluids}, 13(2):105--116, 1992.

\bibitem{Rao1945}
C~Radhakrishna Rao.
\newblock Information and accuracy attainable in the estimation of statistical
  parameters.
\newblock {\em Bull Calcutta. Math. Soc.}, 37:81--91, 1945.

\bibitem{Rogers2007}
Salman~S Rogers, Thomas~A Waigh, Xiubo Zhao, and Jian~R Lu.
\newblock Precise particle tracking against a complicated background:
  polynomial fitting with gaussian weight.
\newblock {\em Physical Biology}, 4(3):220, 2007.

\bibitem{Transtrum2010}
Mark~K Transtrum, Benjamin~B Machta, and James~P Sethna.
\newblock Why are nonlinear fits to data so challenging?
\newblock {\em Physical review letters}, 104(6):060201, 2010.

\bibitem{Zhang2007}
B.~Zhang, J.~Zerubia, and J-C. Olivo-Marin.
\newblock Gaussian approximations of fluorescence microscope point-spread
  functions.
\newblock {\em Appl. Optics}, 46(10):1819--1829, 2007.

\end{thebibliography}

\end{document}